\documentclass[%
 reprint,
superscriptaddress,
 amsmath,amssymb,
 aps,
]{revtex4-2}

\usepackage{graphicx}
\usepackage{dcolumn}
\usepackage{bm}
\usepackage{color}

\begin{document}

\preprint{APS/123-QED}

\title{Active Loop Extrusion guides DNA-Protein Condensation}

\author{Ryota Takaki}
\affiliation{%
 Max Planck Institute for the Physics of Complex Systems 
}%
\author{Yahor Savich}%
\affiliation{%
 Max Planck Institute for the Physics of Complex Systems 
}%
\affiliation{%
 Max Planck Institute of Molecular Cell Biology and Genetics (MPI-CBG), Dresden, Germany
}%
\affiliation{Center for Systems Biology Dresden, Dresden, Germany}
\author{Jan Brugués}%
\email{jan.brugues@tu-dresden.de}
\affiliation{%
 Max Planck Institute for the Physics of Complex Systems 
}%
\affiliation{%
 Max Planck Institute of Molecular Cell Biology and Genetics (MPI-CBG), Dresden, Germany
}%
\affiliation{Center for Systems Biology Dresden, Dresden, Germany}
\affiliation{Cluster of Excellence Physics of Life, TU Dresden, Dresden, Germany}

\author{Frank Jülicher}%
\email{julicher@pks.mpg.de}
\affiliation{%
 Max Planck Institute for the Physics of Complex Systems 
}%
\affiliation{Center for Systems Biology Dresden, Dresden, Germany}
\affiliation{Cluster of Excellence Physics of Life, TU Dresden, Dresden, Germany}




\date{\today}

\begin{abstract}
The spatial organization of DNA involves DNA loop extrusion and the formation of protein-DNA condensates. While the significance of each process is increasingly recognized, their interplay remains unexplored. Using molecular dynamics simulation and theory we investigate this interplay. Our findings reveal that loop extrusion can  enhance the dynamics of condensation and  promotes coalescence and ripening of condensates. Further, the DNA loop enables condensate formation under DNA tension and position condensates. The concurrent presence of loop extrusion and condensate formation results in the formation of distinct domains similar to TADs, an outcome not achieved by either process alone.

\end{abstract}

\maketitle

How cells read and process genomic information represents a fundamental question that is not fully understood. This process involves physical interactions between DNA and proteins that transduce sequence information on DNA to express genes and to organize chromatin. 
Loop extrusion  by structural maintenance of chromosomes (SMC) complexes have been identified as a primary candidate of genome organisation and regulation~\cite{nasmyth2009cohesin,kim2023looping,davidson2021genome,yatskevich2019organization,hoencamp2023genome, polovnikov2023crumpled}. Loop extrusion has been studied through both {\it in vitro} experiments~\cite{terakawa2017condensin,ganji2018real,golfier2020cohesin,kim2022condensin,kim2020dna,pradhan2023smc5,davidson2019dna,kim2019human,ryu2022condensin,pradhan2022smc,ryu2020condensin,shi2018interphase} and theoretical approaches~\cite{takaki2021theory,banigan2019limits,banigan2020loop,banigan2020interplay,chan2023theory,marko2019dna,bonato2021three,chan2024activity}. 
DNA loops are involved in the formation of Topologically Associating Domains (TADs) in chromatin. TAD boundaries are determined by the position of CCCTC-binding factor (CTCF) molecules on the DNA~\cite{yatskevich2019organization,davidson2023ctcf,jeong2024structural,schwarzer2017two,fudenberg2016formation,banigan2023transcription,dey2023structural}.
Another key process involved in chromatin organization is the formation of biological condensates, a process similar to phase separation~\cite{hyman2014liquid,rippe2022liquid,KING20211139,SHRINIVAS2019549,wei2020nucleated,henninger2021rna,morin2022sequence,renger2022co,sommer2022polymer}. Such condensates have been suggested, for example, to play a role in bringing promoters and enhancers into physical proximity~\cite{panigrahi2021mechanisms}.  Indeed condensates have been shown to exert capillary forces which could be involved in such processes~\cite{quail2021force}. These capillary forces are of similar magnitude as forces exerted by SMC molecules during loop extrusion~\cite{quail2021force}. This raises the question of how loop extrusion and condensate formation synergize to organize chromatin.



In this letter, we use simulation and theory to explore the interplay between loop extrusion and protein condensation in the spacial organization of DNA. We report that DNA loops play a pivotal role in nucleating and positioning protein-DNA co-condensates.  The DNA loops not only facilitate the formation of co-condensates but also contribute to their stability under mechanical tension along DNA. We further discuss how loop extrusion and condensation contribute to the emergence of domains in chromatin contact maps, which characterize the DNA spatial organization. 

\begin{figure}[]
\begin{center}
\includegraphics[width=0.45\textwidth]{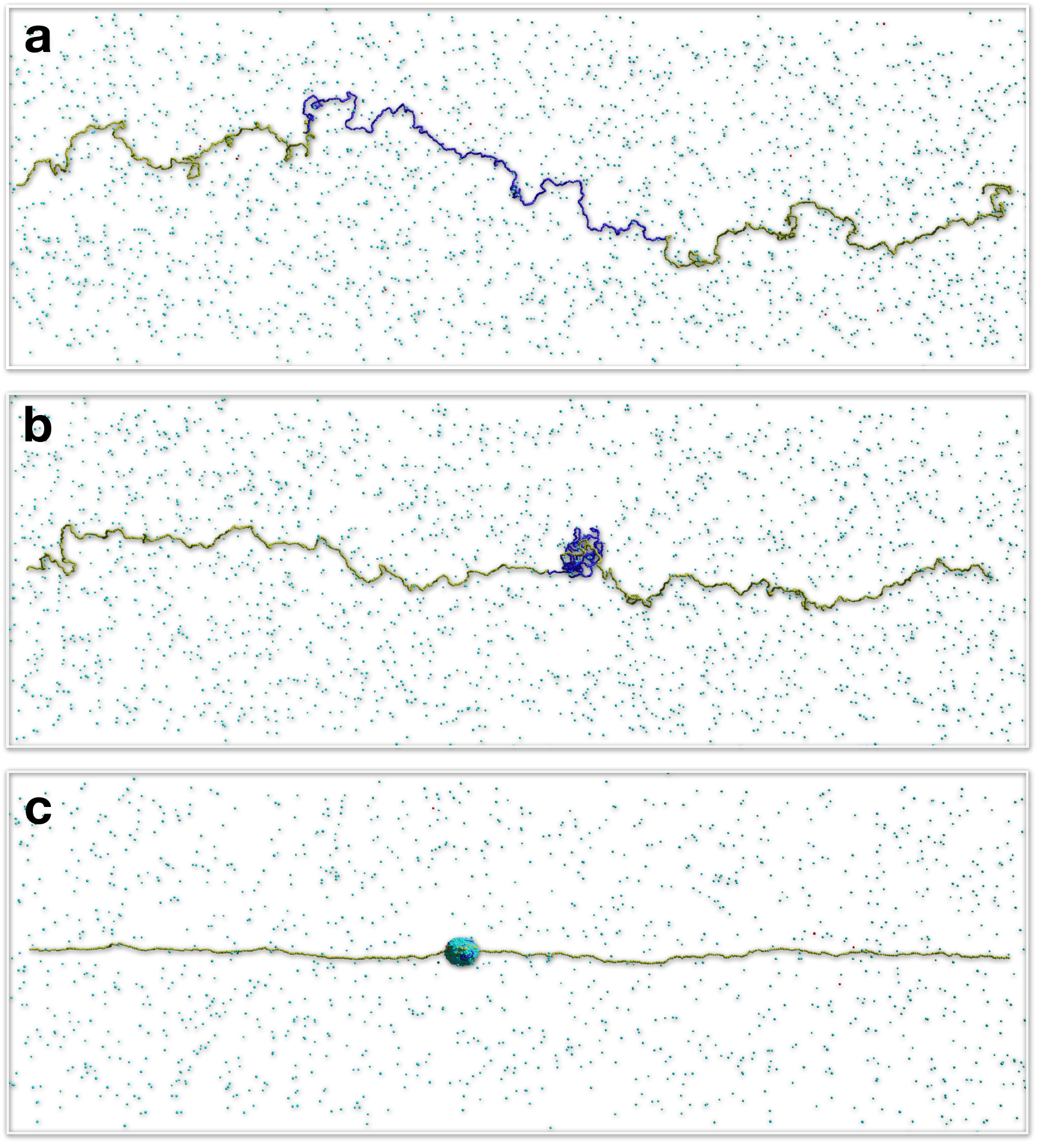}
\end{center}
\caption{\label{Fig:snap}
(a) Initial configuration of the simulation with yellow and blue DNA particles indicating the DNA strand and SMC binding sites on DNA, respectively. Light blue particles represent proteins, and red particles SMC molecules. (b) Example of a DNA loop created by an SMC molecule. (c) Example of a Protein-DNA co-condensate. (a)-(c) are independent simulations for $L_{\mathrm{end}}=6 \ \mathrm{\mu m}$.  }
\end{figure}

Here we consider a configuration often used in  biophysical studies of DNA, where a single DNA molecule is attached at both ends to a  surface ~\cite{ganji2018real,quail2021force}  (Fig.\ref{Fig:snap}). We perform Langevin dynamics simulations that incorporate three distinct types of particle representing DNA segments, proteins, and SMC molecules. These particles interact according to Lennard-Jones potentials as well as FENE potentials along the DNA contour~\cite{rubinsten2003polymer}. DNA-Protein and Protein-Protein interactions are attractive while the interaction among DNA segments is repulsive. 
DNA segments are coarse-grained by particles with $10 \; \mathrm{nm}$ diameter, 
and all three particle types have the same diameter. SMC molecules can bind to the DNA strand in the region indicated in blue in Fig.\ref{Fig:snap}. Upon binding of an SMC molecule to a DNA segment, a two-sided loop extrusion process is initiated. The loop extrusion process terminates when the extruded DNA reaches a boundary of the blue region, mimicking the role of CTCF molecules~\cite{davidson2023ctcf,gabriele2022dynamics}. We start our simulation from randomized initial positions of proteins and DNA segments, with DNA ends fixed at a prescribed distance $L_{\mathrm{end}}\ \mathrm{\mu m}$. The contour length of DNA is set to $L_c=16.5 \ \mathrm{\mu m}$, corresponding to $\lambda$-DNA ~\cite{quail2021force}.  See Supplementary Information (SI) for simulation details. 


\begin{figure}[]
\begin{center}
\includegraphics[width=0.47\textwidth]{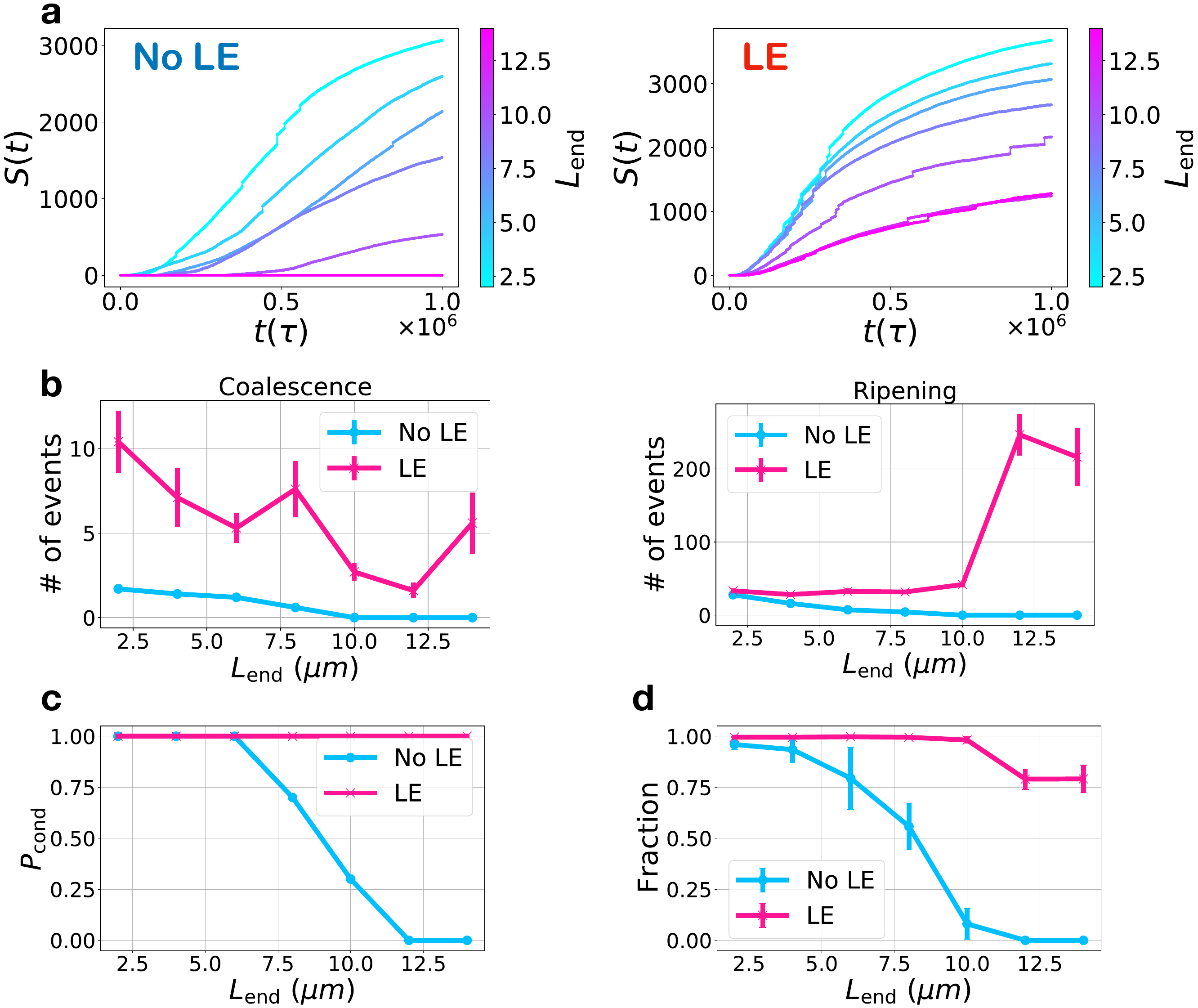}
\end{center}
\caption{\label{Fig:simu} 
(a) Size of the largest condensates $S(t)$  as a function of time $t$. Here $\tau$ is the unit-time in our simulation (SI). The color bars represent the DNA end-to-end distance, $L_{\rm{end}}$. Results without loop extrusion (No LE) and with loop extrusion (LE) are shown.  
(b) Number of condensate disappearance events due to droplet coalescence (left) or disassembly during ripening (right). Lines represent scenarios without loop extrusion (blue) and with loop extrusion (red).
(c) Probability of condensate presence in the final frame of the simulation. (d) Fraction of SMC-binding DNA segments (blue in Fig.\ref{Fig:snap}) within the condensate. 
Error bars indicate the standard error across ten simulations with same parameter values.  }
\end{figure}

We first focus on the dynamics of condensate growth. Similar to conventional droplet kinetics, DNA-protein condensates grow through coalescence and Ostwald ripening when multiple condensates exist. Fig.\ref{Fig:simu}a displays the size $S$ of the largest droplet as a function of simulation time $t$, averaged over ten simulation trajectories. The size $S$ is defined as the number of DNA and protein particles contained in the condensate (section V.A in the SI). Step-wise increases of  $S(t)$ indicate coalescence events, while gradual growth corresponds to ripening.  We compare simulations without loop extrusion (Fig.\ref{Fig:simu}a, left)  to simulations with loop extrusion (Fig.\ref{Fig:simu}a, right) for different $L_{\mathrm{end}}$.  This comparison reveals that loop extrusion accelerates ripening but also enhances coalescence. 

To provide further insight into the condensate growth dynamics, we count the number of events where condensates disappear either by coalescence or by Ostwald ripening, see Fig.\ref{Fig:simu}b. We find a systematic increase of condensate coalescence events in the presence of loop extrusion as compared to the absence of loop extrusion. Furthermore we observe that droplet disappearance by ripening is strongly enhanced for large $L_{\rm{end}}$. The enhancement of Ostwalt ripening by loop extrusion can be understood as follows: condensates outside the loop are subjected to the tension of the polymer and therefore disfavored as compared to the condensate inside the loop which is not subject to tension (section II in the SI).

We next calculated the probability of condensate formation ($P_{\rm cond}$), defined as the probability of condensate formation in the final frame of our simulation trajectories (Fig. \ref{Fig:simu}c). For short $L_{\rm end}$, a condensate is present irrespective of whether a loop was extruded. For larger $L_{\rm end}$ the probability to find a condensate drops and eventually vanishes if no loop is extruded, similar to previous results ~\cite{quail2021force}. Interestingly, loop extrusion enables condensate formation even at large $L_{\rm end}$.  
Loop extrusion can also position condensates. In fact, in our simulations, condensates are often found where the loop is formed.  To quantify this co-localization, we compute in Fig.\ref{Fig:simu}d the fraction of SMC-binding DNA segments (blue in Fig.\ref{Fig:snap}) within the condensate. With loop extrusion, these segments are largely included inside the condensate, even  for larger values of $L_{\mathrm{end}}$. In contrast,  without loop extrusion, this fraction decreases with increasing $L_{\mathrm{end}}$. 

\begin{figure}[]
\begin{center}
\includegraphics[width=0.5\textwidth]{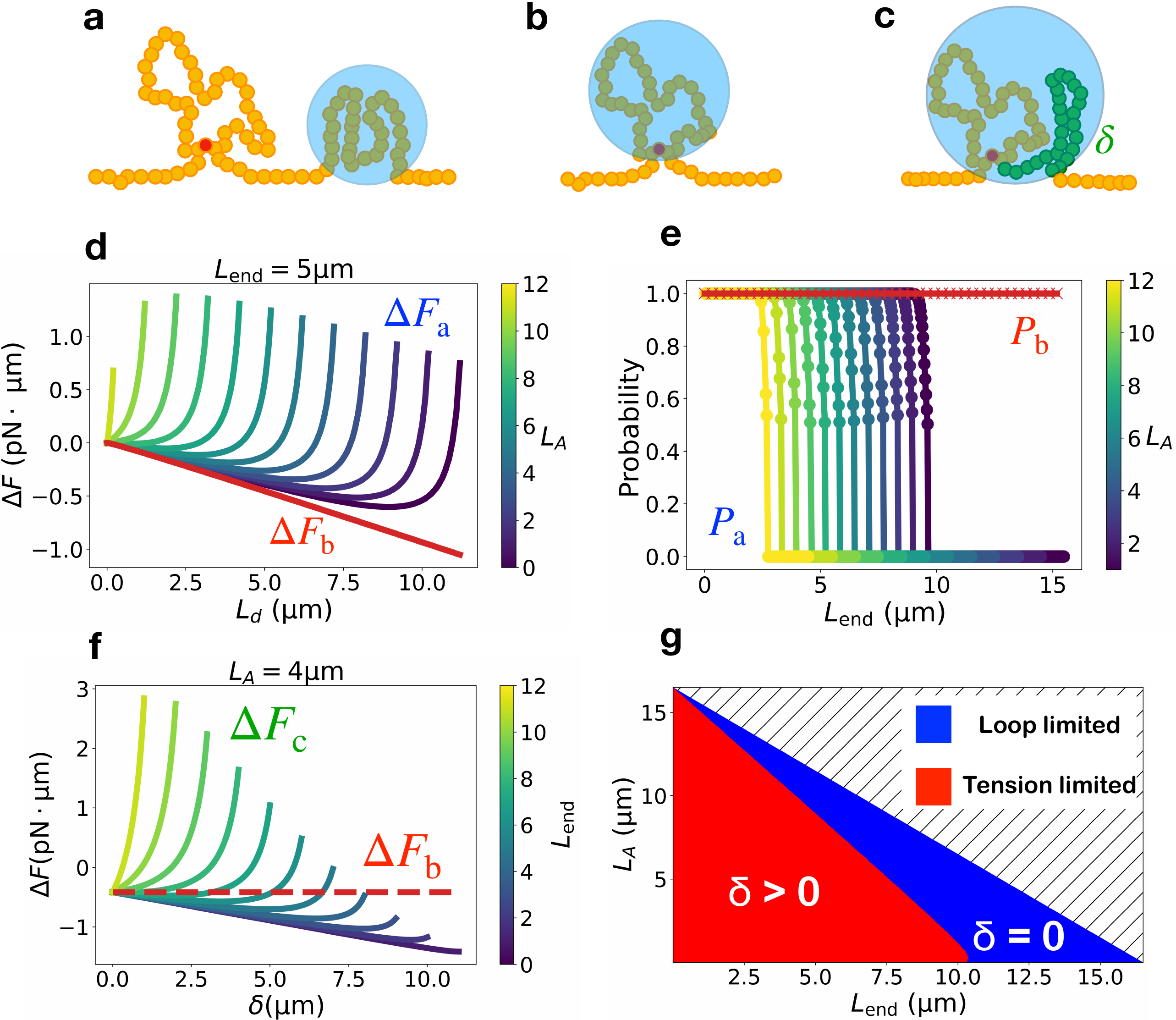}
\end{center}
\caption{\label{Fig:anali} 
(a)-(c) Schematic representations of three scenarios of protein-DNA co-condensation.  
Yellow, blue, and red circle indicates DNA segments, condensate, and SMC protein, respectively.
(a) Condensate positioned outside the DNA loop.  (b) Condensates at the DNA loop. (c) Condensate at the loop, including extra DNA length $\delta$ (green). (d)  Free energy, $\Delta F_a$, for scenario (a) compared to $\Delta F_b$ corresponding to (b), as a function of the length $L_d$ within the condensate for different loop length $L_A$.  
(e) Probability $P_a$ ($P_b$) of condensation formation for scenario (a) (scenario (b)) as a function of $L_{\rm end}$. 
(f) Free energy $\Delta F_c$ for scenario (c) as a function of $\delta$ compared to $\Delta F_b$ for different $L_{\rm end}$. 
(g) Phase diagram of condensation in the presence of DNA loop as a function of $L_A$ and $L_{\rm end}$. The loop limited regime (blue) and the tension limited regime (red) are indicated.  The hatched region is physically inaccessible. 
}
\end{figure}

The positioning of condensates by loop extrusion can be discussed 
by considering three possible scenarios. In scenario (a), the DNA loop is located outside of the condensate (Fig.\ref{Fig:anali}a). In the scenario (b), the condensate is located at the DNA loop and only the DNA loop is inside the condensate (Fig.\ref{Fig:anali}b). Finally in scenario (c) the DNA loop is inside the condensate together with additional DNA segments of length $\delta$ (Fig.\ref{Fig:anali}c).  
Our simulations suggest that condensation  within the DNA loop is not affected by mechanical tension. Therefore condensates form reliably in the scenario (b) and (c) containing the loop.  
To understand the effect of DNA loop on the formation of condensates, we use a simple model of co-condensation~\cite{quail2021force}.
In this model, the free energy of the configuration is the sum, $F=F_d+F_p$, where
\begin{align}
\label{eq:Fd}
F_d(L_d) = -v \alpha L_d + \gamma 4 \pi
\Big(\frac{3 \alpha}{4 \pi}\Big)^{2/3}L_d^{2/3},
\end{align}
is the free energy of the condensate 
and
\begin{align}
\begin{split}
\label{eq:Fp}
\small
F_p(L_d,L_{\mathrm{end}},L_c) &= \frac{k_BT L_{\mathrm{end}}^2}{4 l_p}\Big(\frac 1{L_c-L_d-L_{\mathrm{end}}}+\frac 2{L_c-L_d}\Big)
\end{split}
\end{align}
is the free energy of the non-condensed DNA. Here $L_d$ is the length of the DNA segments inside the condensate, $k_B$ is the Boltzmann constant, $T$ represents the temperature, and $l_p$ is the persistence length of DNA. The parameters $\alpha$, $\gamma$, and $v$ are the inverse of the DNA packing density, surface tension of the condensate, and condensation free energy per volume, respectively \cite{quail2021force}. 
Below we use parameters values obtained for forkhead box protein A1~\cite{quail2021force}: $\alpha=0.04 \ \mathrm{\mu m^2}$, $\gamma=0.04 \ \mathrm{pN\:\mu m ^{-1}}$, $v=2.6 \ \mathrm{pN\:\mu m ^{-2}}$ and $l_p=50$ nm.

We first consider scenario (a) (Fig.\ref{Fig:anali}a). In this case the condensate is located outside the DNA loop of length $L_A$. We define the energy difference $\Delta F_{a} = F_d(L_d) + F_p(L_d,L_{\rm end}, L_{\mathrm{c}}-L_A)-F_p(0,L_{\rm end}, L_{\mathrm{c}}-L_A)$ as the difference of the free energy before and after condensate formation, where we have taken into account that the DNA contour length $L_c$ is effectively reduced by the loop length $L_A$. 
The stable condensate size is then obtained by minimizing $\Delta F_a$ with respect to $L_d$. 

In scenario (b), the condensate contains the DNA length $L_d=L_A$ and the free energy change due to condensate formation is simply given by $\Delta F_b=F_d(L_A)$.
Correspondingly, in scenario (c), we define $\Delta F_c=F_d(L_A +\delta) + F_p(L_A + \delta,L_{\rm end}, L_{\mathrm{c}}) - F_p(L_A,L_{\rm end}
, L_{\mathrm{c}})$. When $\delta=0$, the DNA length inside the condensate equals to the loop length $L_A$, reducing this case to scenario (b). 



We show $\Delta F_{a}$ in Fig.\ref{Fig:anali}d, as well as $\Delta F_{b}$ for $L_{\mathrm{end}}=5 \ \mathrm{\mu m}$.
Increasing the loop length $L_A$ shifts the minimum position of $\Delta F_{a}$ towards smaller values of $L_d$ until the condensate vanishes via a first-order phase transition, similar to the one reported previously~\cite{quail2021force}. 
Fig.\ref{Fig:anali}d reveals that $\Delta F_b <\Delta F_a$, implying that a condensate will always form inside the loop, irrespective of the tension on the DNA. Thus, the DNA loop guides condensate formation. Fig.\ref{Fig:anali}e, shows the conditional probability of generating a condensate outside the loop ($P_{
a}$) as a function of $L_{\mathrm{end}}$. 
Increasing $L_A$ shifts the curves for $P_{
a}$ towards smaller $L_{\mathrm{end}}$ values: the tension induced by the DNA loop narrows the range where condensation outside the loop is possible. 
However condensates in the loop are always favored ($P_b=1$). 


Fig.\ref{Fig:anali}f shows the free energy profile $\Delta F_c$ as a function of $\delta$ together with $\Delta F_b$. 
This shows that for sufficiently large $L_{\rm end}$ the free energy is minimal for $\delta =0$ and the condensate size is equal to the loop size. We call this loop limited condensate. As $L_{\rm end}$ is decreased below the critical value $L_{\rm end}^c$, $F_c$ exhibits a minimum at $\delta >0$ corresponding to a condensate containing a DNA segment that is longer than the loop. In this regime the condensate is tension limited. 
At $L_{\rm end}=L^c_{\rm end}$ where $\delta$ vanishes,  a continuous second order transition occurs. 
Fig.\ref{Fig:anali}g shows the phase diagram for condensates in the presence of a loop. In the blue region, $\delta=0$ and condensates are loop limited.  In the red region, $\delta > 0$ and condensates are tension limited. Both regions meet at a second order transition line. 
 
We find that the phase transition line between the loop limited and tension limited regimes lies within the force range where CTCF molecules efficiently regulate the loop extrusion process of cohesin~\cite{davidson2023ctcf}. This suggests that the loop limited regime may play a role in regulating TAD formation (see Section III in the SI for further analysis).

\begin{figure}[]
\begin{center}
\includegraphics[width=0.5\textwidth]{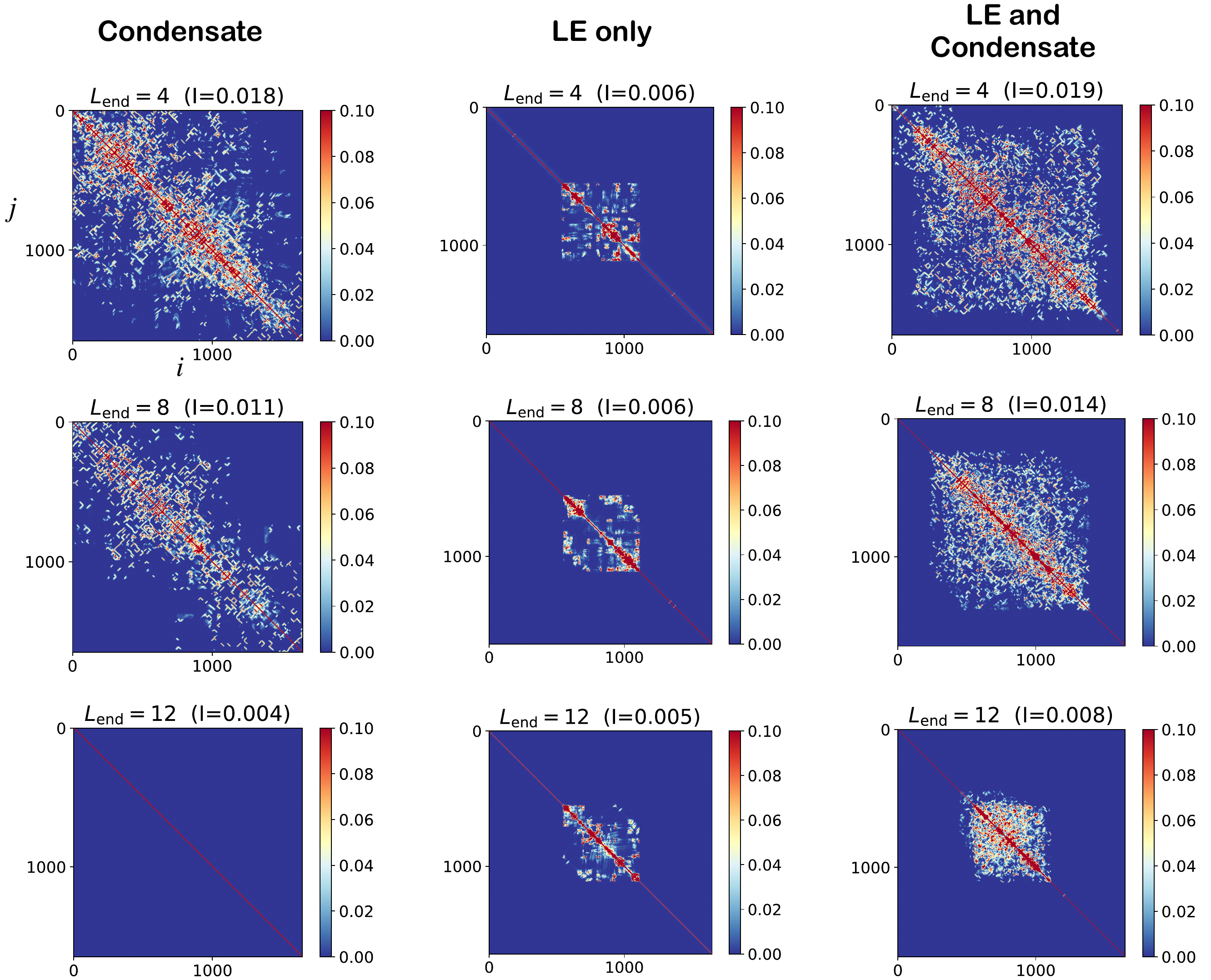}
\end{center}
\caption{\label{Fig:contact} Contact maps obtained in simulations for only condensation (left), only loop extrusion (middle), and both loop extrusion and condensation (right) as a function of DNA particle indices $i$ and $j$. Contact maps are presented for $L_{\mathrm{end}}=4,8,12 \ \mathrm{\mu m}$ (top to bottom).
The contact probability between particle $i$ and $j$ is shown as a color code. The fraction $I$ of contacts,  defined as the sum of the all contact probabilities normalized by the total number of pairs ($1650^2$), is given.  }
\end{figure}

We now discuss the effects of loops on DNA conformations inside the condensates. Both condensation and loop extrusion lead to a local accumulation of DNA segments, see Fig. \ref{Fig:snap}. Such accumulation fosters contacts between DNA segments, even when they are at a distance along the sequence. The probabilities of such contacts are characterized by a contact map, a key tool to understand chromatin organization~\cite{lieberman2009comprehensive, shi2019conformational,shi2023maximum,shi2021hi,shin2023effective}.  We determine contact maps for different end-to-end distance in our simulations. 
Fig.\ref{Fig:contact} presents examples of contact maps with contact probability between two segments $i$ and $j$, where $i$ and $j$ are the DNA particle indices of two DNA segments. 
In Fig. \ref{Fig:contact}, contact maps are shown for DNA-protein co-condensation without loop extrusion (left column),  for loop extrusion without condensation (middle column), and for both condensation and loop extrusion (right column), for different values of $L_{\rm end}$.
We find that if only condensation happens, contact maps show many but irregularly positioned disordered contacts 
for short $L_{\rm end}$. These contacts disappear as $L_{\rm end}$ is increased and condensates dissolve. For only loop extrusion, contacts occur within the loop and are dominated by short-ranged contacts. When condensation and loop extrusion are combined, square patterns of contacts resembling TADs emerge. These square patterns imply that contacts over longer distances along the chain are prominent. 

\begin{figure}[]
\begin{center}
\includegraphics[width=0.5\textwidth]{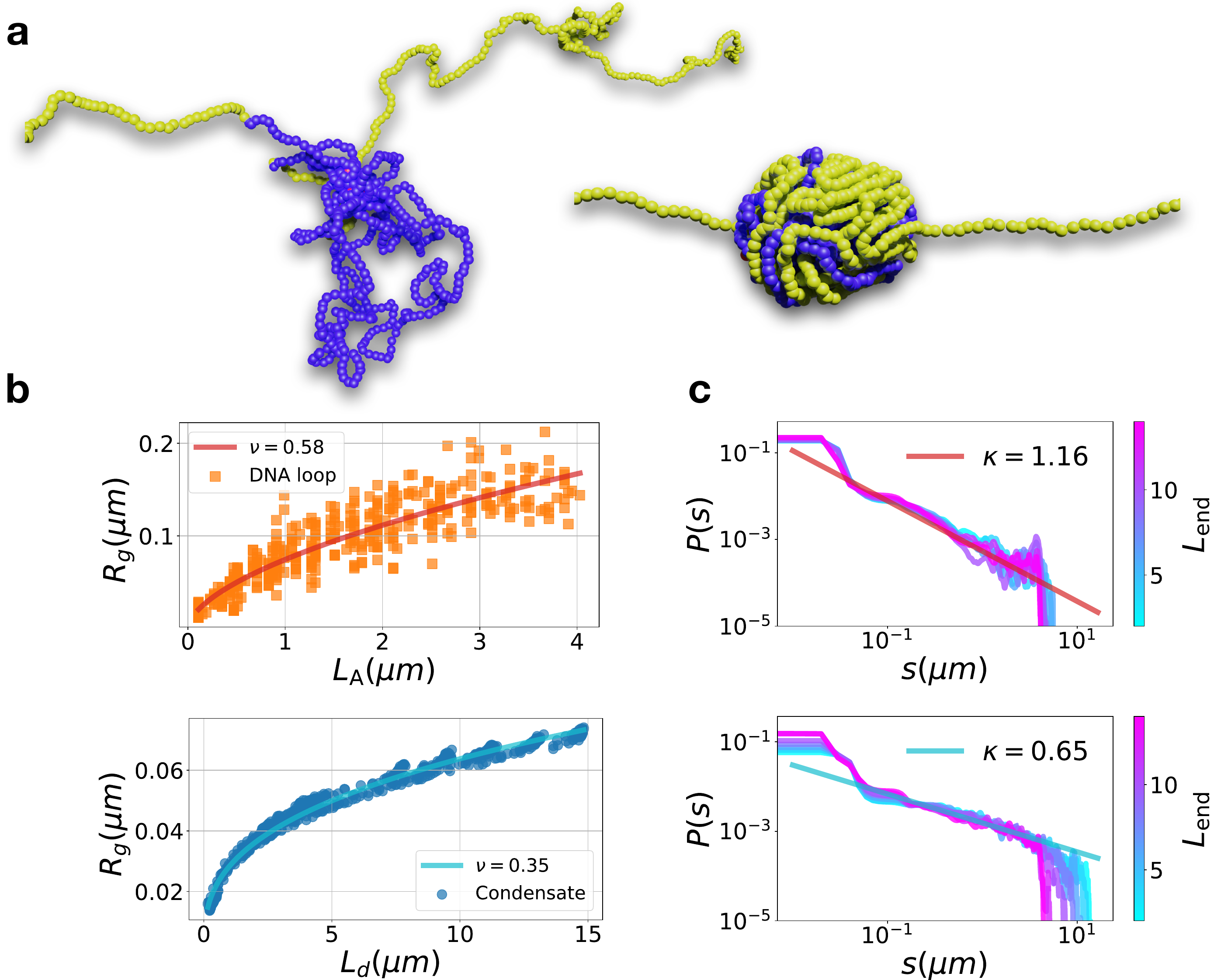}
\end{center}
\caption{\label{Fig:blobs} 
(a) Representative snapshots showing a DNA loop without condensation (left) and a DNA-Protein condensate containing the DNA loop (right).  Yellow and blue particles indicate the DNA strand and SMC binding site on DNA, respectively. Protein particles are not shown to emphasize the DNA structure. $L_{\mathrm{end}}=6\ \mu m$.
(b) Radius of gyration $R_g$ of DNA loops without condensation as a function loop length $L_A$ (top, $\nu \simeq 0.58$) and of DNA within a condensate as a function of condensed DNA length $L_d$ (bottom,  $\nu \simeq 0.35$). The data is fit using $R_g \sim L^{\nu}$ (solid lines). 
(c) Probability of contact inside loops and condensates, $P$, as a function of the DNA length between contacts, $s$, for different $L_{\rm end}$. Loop extrusion without condensation (top, $\kappa=1.16$) and loop extrusion with condensation (bottom, $\kappa=0.65$). The data is fit using $P \sim s^{-\kappa}$ (solid lines).
The reported values of $\kappa$ are the mean value of the exponents for different $L_{\rm end}$  fit in the range $5\cdot 10^{-2}\ \mu m < s < 3\ \mu m$.
}
\end{figure}

To further characterise the structures generated by condensation and loop extrusion, we consider the polymeric configuration of loops and condensates.
Fig.\ref{Fig:blobs}a shows example configurations obtained in simulations: a DNA loop without condensation (left)  and a DNA loop within a condensate (right).  This reveals a coil-like structure of the DNA loop and a densely packed DNA within the condensate. Fig.\ref{Fig:blobs}b shows,  the radius of gyration ($R_g$) of the DNA loop as a function of loop length, exhibiting a scaling behavior with an exponent $\nu=0.58$, similar to the Flory exponent $\nu=3/5$ of a polymer in a good solvent~\cite{rubinsten2003polymer}. In contrast,  $R_g$ of the DNA inside a condensate exhibits different scaling behavior with $\nu=0.35$, which is close to the exponent $\nu=1/3$ of a collapsed polymer in a poor solvent~\cite{rubinsten2003polymer}. 
Using these results,  we can explain the difference between the contact maps shown in Fig.\ref{Fig:contact}. 
In the absence of condensation,  the looped polymer behaves as a random coil favoring short range contacts. Indeed, the contact probability decreases for increasing distance along the polymer, giving rise to the short ranged contact maps shown in Fig.\ref{Fig:contact}, middle.   
For a loop inside the condensate, the contact probability remains high, giving rise to the longer ranged contact map.

In Fig.5c, we plot the contact probability density $P(s)$ as a function of the DNA length $s$ between contact points. We find that the probability density follows a power law, $P \sim s^{-\kappa}$. In the scenario with only loop extrusion, we observe $\kappa \simeq 1.16$ (Fig.5c, top). However, when condensate formation occurs in addition to loop extrusion, the exponent decreases significantly to $\kappa = 0.65$ (Fig.5c, bottom). This result is surprising, as the commonly used mean-field argument, $\kappa = 3\nu$~\cite{halverson2014melt}, suggests $\kappa \geq 1$ for $\nu \geq 1/3$. Interestingly, an exponent $\kappa \sim 0.7$ has been observed in experiments~\cite{schwarzer2017two,sanborn2015chromatin}, which is attributed to the scaling regime related to TADs. 
Such exponent has been suggested to rise because of a non-equilibrated structure of extruded loops~\cite{chan2024activity} and exponentially distributed loops~\cite{polovnikov2023topological,polovnikov2023crumpled}.  Our findings show that long-range DNA contacts mediated by protein-DNA co-condensation could also yield an exponent smaller than 1.

In summary, we investigated the interplay between DNA loop extrusion and DNA-protein co-condensation. We found that loop extrusion stabilises DNA-protein co-condensation under tension and positions the condensate on DNA.
We identified a regime of loop limited condensates, where condensates size is set by the DNA loop size. 
Our work shows that by combining loop extrusion and condensation,  a DNA organization with the characteristics of TADs can naturally emerge. 
In the absence of condensation, the resulting contact maps remain short ranged.
The condensation facilitates close contacts of distant DNA segments within the TAD. 
In our work, we use simple simulation assay that allows us to control tension along DNA. While we do not recapitulate the complexities of chromatin in the cell nucleus, we think that this assay reveals physical principles that could play a key role for genome organization in the cell. 
Our work therefore underscores that chromatin organization in TADs may emerge from the interplay between loop extrusion and protein-DNA co-condensation.

\section*{Acknowledgements}
F. J. acknowledges funding from the Volkswagen Foundation. R.T. thanks to the overseas research fellowship No.
202260312 from the Japan Society for the Promotion of Science.

\bibliography{mybib}

\end{document}


\pagebreak
\widetext
\begin{center}
\textbf{\large Supplementary Information: Active Loop Extrusion guides DNA-Protein Condensation}
\end{center}

\author{Ryota Takaki}
\affiliation{%
 Max Planck Institute for the Physics of Complex Systems 
}%
\author{Yahor Savich}%
\affiliation{%
 Max Planck Institute for the Physics of Complex Systems 
}%
\affiliation{%
 Max Planck Institute of Molecular Cell Biology and Genetics (MPI-CBG), Dresden, Germany
}%
\author{Jan Brugués}%
\email{jan.brugues@tu-dresden.de}
\affiliation{%
 Max Planck Institute for the Physics of Complex Systems 
}%
\affiliation{%
 Max Planck Institute of Molecular Cell Biology and Genetics (MPI-CBG), Dresden, Germany
}%
\affiliation{Center for Systems Biology Dresden, Dresden, Germany}
\affiliation{Cluster of Excellence Physics of Life, TU Dresden, Dresden, Germany}

\author{Frank Jülicher}%
\email{julicher@pks.mpg.de}
\affiliation{%
 Max Planck Institute for the Physics of Complex Systems 
}%
\affiliation{Center for Systems Biology Dresden, Dresden, Germany}
\affiliation{Cluster of Excellence Physics of Life, TU Dresden, Dresden, Germany}

\setcounter{equation}{0}
\setcounter{figure}{0}
\setcounter{table}{0}
\setcounter{page}{1}
\makeatletter
\renewcommand{\theequation}{S\arabic{equation}}
\renewcommand{\thefigure}{S\arabic{figure}}
\renewcommand{\bibnumfmt}[1]{[S#1]}
\renewcommand{\citenumfont}[1]{S#1}
\maketitle

\section{Langevin dyanmics simulations}
Our simulation is comprised of three types of particle: 1650 DNA particles, 5000 protein particles, and 20 SMC particles. 
The equation of motion of the particles is given by Langevin equation, 
\begin{align}
m \frac{d \vb{v}_i}{dt} = -\nabla U  - m\gamma \vb{v}_i + \vb{\xi}. 
\end{align}
Here $m$ is the mass of the particles, $U$ is the potential energy, $\gamma$ is the friction coefficient (inverse time unit), $\vb{v}_i$ is the velocity of $i_{th}$ particle, and $\xi$ is the delta correlated noise satisfying the fluctuation dissipation relation. 

The potential energy $U$ is given by the contributions: $U=U_{\mathrm{LJ}}+U_{\mathrm{FENE}}+U_{\mathrm{ANG}}$, where $U_{\mathrm{LJ}}$ (Lennard-Jones potential), $U_{\mathrm{FENE}}$ (Finitely Extensible Nonlinear Elastic potential), and $U_{\mathrm{ANG}}$ (angular potential).
We employ the Lennard-Jones potential to simulate attraction and soft repulsion interactions between DNA-Protein, Protein-Protein, and DNA-SMC. The potential is defined as:

\begin{equation}
U_{\mathrm{LJ}}(r_{i,j}) = 4\epsilon_{\alpha_i,\alpha_j} \Big( \big(\frac{\sigma}{r_{i,j}}\big)^{12}-\Delta_{\alpha_i,\alpha_j} \big(\frac{\sigma}{r_{i,j}}\big)^{6} \Big)
\Theta(r_c-r_{i,j}).
\end{equation}

In this expression, $\epsilon$ and $\sigma$ denote the strength and range of the interaction, respectively. The indices $\alpha_i$ and $\alpha_j$ specify the types of particles (DNA, Protein, or SMC) for particles $i$ and $j$. The function $\Theta$ is the Heaviside step function, introducing a cut-off at $r_c=5 \sigma$. The strength and combination among different particle types in the Lennard-Jone potential are determined by the matrix $\epsilon$ and $\Delta$, respectively.

\begin{align}  
\begin{split}
(\bm{\epsilon}_{\alpha_i \alpha_j})=
\begin{pmatrix}
\small
\epsilon_{\mathrm{DNA},\mathrm{DNA}} & \epsilon_{\mathrm{DNA},\mathrm{PRO}} & \epsilon_{\mathrm{DNA},\mathrm{SMC}}\\
\epsilon_{\mathrm{DNA},\mathrm{PRO}} & \epsilon_{\mathrm{PRO},\mathrm{PRO}} & \epsilon_{\mathrm{PRO},\mathrm{SMC}}\\
\epsilon_{\mathrm{DNA},\mathrm{SMC}} & \epsilon_{\mathrm{PRO},\mathrm{SMC}} & \epsilon_{\mathrm{SMC},\mathrm{SMC}}\\
\end{pmatrix}
\end{split}  
\end{align}

\begin{equation} 
(\bm{\Delta}_{\alpha_i \alpha_j})=
\begin{pmatrix}
0 & 1 & 1\\
1 & 1 & 0\\
1 & 0 & 0\\
\end{pmatrix}
\end{equation}

To maintain the connectivity of DNA particles, we utilize the Finitely Extensible Nonlinear Elastic potential:
\begin{align}
U_{\mathrm{FENE}}(r_{i,i+1}) = -\frac{1}{2}k_F R_F^2 \log\Big[1-\frac{(r_{i,i+1}-r^0_{i,i+1})^2}{R_F^2}\Big],
\end{align}
where $k_F$ is the stiffness of the potential, $R_F$ denotes the maximum allowable displacement, and $r^0_{i,i+1}$ is the equilibrium distance between adjacent particles $i$ and $i+1$, set to $r^0_{i,i+1}=\sigma$. We set $k_F=30\epsilon/\sigma^2$ and $R_F=1.5 \sigma$.

To accurately model the stiffness of DNA, we apply the angular potential:
\begin{equation}
U_{\mathrm{ANG}}(\theta_{k}) =\epsilon_b(1+\cos \theta_k),
\end{equation}
where $\epsilon_b$ is the energy scale for bending. It is related to the polymer's persistence length by $\epsilon_b=k_BT l_p/\sigma$~\cite{midya2019phase}.

{\bf Modeling loop extrusion}.
To simulate the non-equilibrium, active loop extrusion (LE) process, we initiate a DNA loop upon the binding of an SMC particle to a DNA particle. This binding and subsequent loop formation are governed by the activation and deactivation of breakable harmonic potentials.
The binding potential between SMC and DNA is described as:
\begin{equation}
U_{\mathrm{SMC,DNA}}(r_{i,j}) = -\epsilon_{\mathrm{SMC,DNA}} \log\Big(1+e^{-\frac{1}{2}k_{\mathrm{SMC,DNA}}(r_{i,j}-\sigma)^2}\Big),
\end{equation}
where $\epsilon_{\mathrm{SMC,DNA}}$ and $k_{\mathrm{SMC,DNA}}$ are the strength and stiffness of the potential, respectively.
The loop extrusion potential is given by:
\begin{equation}
U_{\mathrm{LP}}(r_{i,j}) = -\epsilon_{\mathrm{LP}} \log\Big(1+e^{-\frac{1}{2}k_{\mathrm{LP}}(r_{i,j}-\sigma)^2}\Big),
\end{equation}
with $\epsilon_{\mathrm{LP}}$ and $k_{\mathrm{LP}}$ being the corresponding strength and stiffness parameters.

The loop extrusion process implemented as sequence of events, divided in three steps:  1. SMC binding, 2. Loop extrusion 3. End of the loop extrusion. Each of these steps contains three subsequent substeps as follows
\begin{enumerate}
    \item SMC binding to DNA 
    \begin{enumerate}
        \item SMC binds to a DNA particle when the distance between them is less than the cutoff distance of $1.5\sigma$. Assume SMC binds at DNA index $i_{\mathrm{SMC}}$.
        \item Activate the binding potential $U_{\mathrm{SMC,DNA}}$ between SMC and DNA, with parameters $k_{\mathrm{SMC,DNA}}=0.005(k_BT/\nm^2)$ and $\epsilon_{\mathrm{SMC,DNA}}=100(k_BT)$. Deactivate the attractive interaction in $U_{\mathrm{LJ}}$ between SMC and DNA.
        \item Activate the first cross-link potential $U_{\mathrm{LP},1}$ between $i_{\mathrm{SMC}}$ and DNA particle at $i_{\mathrm{SMC}}\pm i_0$ (where $i_0=10$ is the initial loop size), in a randomly chosen loop extrusion direction ($\pm$). Also, activate the second potential $U_{\mathrm{LP},2}$ between $i_{\mathrm{SMC}}$ and $i_{\mathrm{SMC}}\pm i_0\pm1$ in the chosen extrusion direction. Initial parameters for $U_{\mathrm{LP}}$ are $\epsilon_{\mathrm{LP}}=100(k_BT)$ and $k_{\mathrm{LP}}=0.001(k_BT/\nm^2)$.
    \end{enumerate}
    \item Loop extrusion 
        \begin{enumerate}
        \item Move $U_{\mathrm{LP},1}$ to the pair of beads $i_{\mathrm{SMC}}\pm 1$ and $i_{\mathrm{SMC}}\mp i_0 \mp 2$. Simulate for $n_{\mathrm{LE}}/2$ steps, where $1/n_{\mathrm{LE}}$ defines the loop extrusion speed. We set $n_{\rm LE}=10^6$ in order for a maximum steps in the simulation to be $100$ loop extrusion steps ($\sim 4 \mathrm{\mu m}$ DNA loop). Parameters for $U_{\mathrm{LP}}$ remain the same.
        \item Move $U_{\mathrm{LP},2}$ to the pair $i_{\mathrm{SMC}}\pm 2$ and $i_{\mathrm{SMC}}\mp i_0 \mp 3$. Simulate for another $n_{\mathrm{LE}}/2$ steps. Parameters for $U_{\mathrm{LP}}$ are unchanged.
        \item Continue (a) and (b) until SMC unbinds from DNA, the loop breaks, or the loop extrusion reaches the determined end of DNA. 
    \end{enumerate}
    \item End of the loop extrusion 
        \begin{enumerate}
        \item Unbinding of SMC from DNA. If the distance between SMC and SMC bound DNA becomes greater than some threshold distance $d=100 \sigma$, then stop the loop extrusion by removing $U_{\mathrm{SMC,DNA}}$, $U_{\mathrm{LP},1}$ and $U_{\mathrm{LP},2}$.  
        \item Breaks of the loop.  If the distance between DNA particles associated with $U_{\mathrm{LP},1}$ or $U_{\mathrm{LP},2}$ becomes greater than some threshold distance $d=100 \sigma$, then stop the loop extrusion by removing $U_{\mathrm{SMC,DNA}}$, $U_{\mathrm{LP},1}$ and $U_{\mathrm{LP},2}$.   
        
        \item Loop extrusion reaches to the determined end of DNA. Once the loop extrusion reaches to the end position, stop updating the position of $U_{\mathrm{LP},1}$ and $U_{\mathrm{LP},2}$.
    
    \end{enumerate}
\end{enumerate}

In the Langevin dynamics simulation we use the reduced units, 
\begin{equation}
m=\frac{k_BT \tau^2}{\sigma^2}; \quad \tau=1/\gamma,
\end{equation}
where $\tau$ is the unit time in our simulation. 
We performed Langevin dynamics simulations using the openMM software package~\cite{eastman2017openmm} with the time step $\Delta t= 0.01 \tau$.
Simulations were carried out in a rectangular box with dimensions $2 \mathrm{\mu m} \times 2 \mathrm{\mu m} \times 15 \mathrm{\mu m}$, implementing periodic boundary conditions.
{\color{cyan} To reproduce experimentally consistent phase separation behavior, we calibrated $\epsilon_{\rm PRO, PRO} = 2 \ k_BT$, which results in a condensation probability as a function of end-to-end distance similar to experimental findings~\cite{quail2021force}. For smaller values ($\epsilon_{\rm PRO, PRO} \lesssim 1.5\ k_BT$), phase separation does not occur, while for higher values ($\epsilon_{\rm PRO, PRO} \gtrsim 3\ k_BT$), condensates form regardless of the end-to-end distance. For the loop extrusion process, we set the motor to extrude a maximum of $4\ \mu m$ of DNA during the condensation formation (a few seconds in experiments).
}

\begin{table*}[t]
\begin{center}
  \begin{tabular}{|c | c| c|}
    \hline
    \text{Parameter} &Meaning &\text{Value}\\
    \hline \hline
    $\epsilon_{\mathrm{DNA},\mathrm{DNA}}$ & DNA-DNA interaction for LJ potential    &$1 \ k_BT$\\
    $\epsilon_{\mathrm{SMC},\mathrm{SMC}}$ &LEF-LEF interaction for LJ potential&$1 \ k_BT$\\
    $\epsilon_{\mathrm{SMC},\mathrm{DNA}}$ & LEF-DNA interaction for LJ potential    &$10 \ k_BT$\\
    $\epsilon_{\mathrm{PRO},\mathrm{PRO}}$ &Protein-Protein interaction for LJ potential  &2$ \ k_BT$\\
    $\epsilon_{\mathrm{DNA},\mathrm{PRO}}$ &DNA-Protein interaction for LJ potential &Same as $\epsilon_{\mathrm{PRO},\mathrm{PRO}}$ \\
    $\sigma$ & Size of a particle        &$10 \ \mathrm{nm}$\\
    $\epsilon_b$ & Energy scale for angle potential          &$5 \ k_BT$\\
    \hline
  \end{tabular}
\end{center} 
\caption{\label{Table:parameters} List of parameters and the values of the potential chosen in the molecular dynamics simulation.}
\end{table*}

\section{Condensate growth and Ostwalt ripening under DNA tension}
We here discuss the dynamics of condensate growth under the tension on DNA, which leads to the enhanced Ostwalt ripening.  Let us consider the situation in Fig.3a in the main text where a condensate is located outside the DNA loop. We define the difference of the chemical potential ($\Delta \mu$) between the condensate ($\mu_d$) and the stretched DNA ($\mu_p$),  
\begin{align} 
\begin{split}
\label{}
\Delta \mu \equiv \sigma  \frac{\partial \Delta F_a(L_d,L_A, L_{\rm end},L_c)}{\partial L_d}=\mu_d(L_d) - \mu_p (L_d,L_A,L_{\rm end},L_c),
\end{split}  
\end{align}
where 
\begin{align} 
\begin{split}
\label{}
\mu_d(L_d) \equiv    \sigma \frac{\partial F_d(L_d)}{\partial L_d}= \sigma \left( -v \alpha  + \gamma \frac{8 \pi}{3}
\Big(\frac{3 \alpha}{4 \pi}\Big)^{2/3}L_d^{-1/3} \right)
\end{split}  
\end{align}
and 
\begin{align} 
\begin{split}
\label{}
\mu_p (L_d,L_A,L_{\rm end},L_c) \equiv    -\sigma \frac{\partial F_p(L_d, L_{\rm end}, 
 L_c-L_A)}{\partial L_d}= \sigma \left( -\frac{k_BT }{4 l_p}\left(\frac{L^2_{\rm end}}{(L_{\rm end}+L_d+L_A-L_c)^2} + \frac{2L^2_{\rm end}}{(L_d+L_A-L_c)^2} \right) \right).
\end{split}  
\end{align}
Here $\sigma$ is the length of a DNA segment which is set to $1$ for simplicity.
To investigate the dynamics of the condensate growth, let us write,
\begin{align} 
\begin{split}
\label{eq:dot_Ld}
\dot{L}_d =- \gamma \Delta \mu ,
\end{split}  
\end{align}
where $\gamma$ is a motility coefficient. The equilibrium is given by the condition $\Delta \mu=0$.
Eq.\ref{eq:dot_Ld} has two fixed points for $L_d$ corresponding to the two stages of the condensate growth.  First stage is the initial nucleation of protein-DNA condensate and the second stage is the growth of the condensate by further inclusion of DNA into the condensate. The first process is approximately independent to the DNA tension while the second process is tension dependent.

\begin{figure}[h]
\begin{center}
\includegraphics[width=\textwidth]{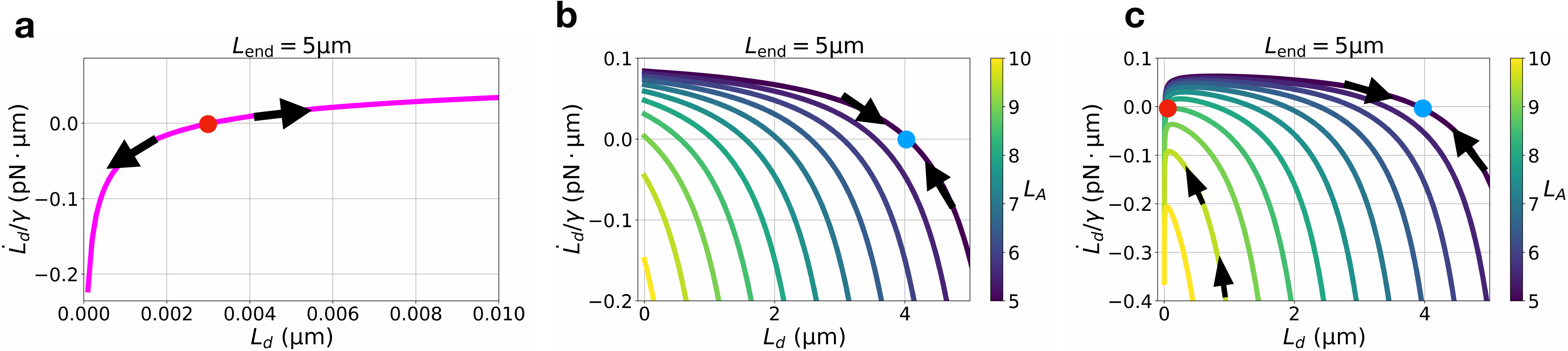}
\end{center}
\caption{\label{Fig:fixed_points} (a) The first fixed point obtained from Eq.\ref{eq:first_FP}. The red dot indicates the unstable fixed point and the arrows are the direction of the evolution of $L_d$. (b) The second fixed point obtained from Eq.\ref{eq:second_FP}. The blue dot indicates an example of stable fixed point for a value of $L_A$. The arrows are the direction of the evolution of $L_d$. The color bar shows the values of the DNA loop length $L_A$ due to the loop extrusion process. (c) The dynamics of the condensate growth from Eq.\ref{eq:dot_Ld}. The red and blue dots are the unstable and stable fixed points, respectively. 
}
\end{figure}

To illustrate the two fixed points, we start by obtaining the positions of the two fixed points. The first fixed point corresponds the initial nucleation of the condensate, which is approximately obtained by 
\begin{align} 
\begin{split}
\label{eq:first_FP}
 v \alpha  - \gamma \frac{8 \pi}{3}
\Big(\frac{3 \alpha}{4 \pi}\Big)^{2/3}L_d^{-1/3}= 0.
\end{split}  
\end{align}
This fixed points is unstable. The condensates smaller than the critical size diminish while the condensates larger than the critical size grow.  See Fig.\ref{Fig:fixed_points}a.

The second fixed point corresponds to the condensate growth by the inclusion of DNA, which is approximately given by 

\begin{align} 
\begin{split}
\label{eq:second_FP}
 v\alpha -\frac{k_BT }{4 l_p}\left(\frac{L^2_{\rm end}}{(L_{\rm end}+L_d+L_A-L_c)^2} + \frac{2L^2_{\rm end}}{(L_d+L_A-L_c)^2} \right) =0.
\end{split}  
\end{align}
This fixed point is stable, see Fig.\ref{Fig:fixed_points}b.

Therefore the dynamics of condensate growth, Eq.\ref{eq:dot_Ld}, has two fixed points, one is unstable and the other is stable (Fig.\ref{Fig:fixed_points}c). The effect of the DNA loop, $L_A$, has significant effect to the dynamics of the condensate growth. High values of $L_A$ lead to the loss of the fixed points rendering any formation of condensates $outside$ the loop unstable (Fig.\ref{Fig:fixed_points}c, see the loss of the intersection at $\dot{L}_d/\gamma =0$ for the high values of $L_A$).  Therefore when the condensate within DNA loop and outside the DNA loop coexist, the condensate outside loop tends to diminish for high $L_A$. In turn, the condensate inside the loop, which is stable, should grow instead. Thus the tension along DNA (DNA loop by loop extrusion process in our study) accelerates the dynamics of Ostwalt ripening of condensate within the DNA loop.

\section{Connection to the tension-dependent regulation of loop extrusion process by CTCF}
\begin{figure}[]
\begin{center}
\includegraphics[width=0.7\textwidth]{force_CTCF_stall_2.pdf}
\end{center}
\caption{\label{Fig:force_CTCF} Comparison of DNA tension in our study with tension regimes derived from Davidson et al.~\cite{davidson2023ctcf}. {\bf Left:} The blue curve represents the tension along the DNA computed using Eq.~\ref{eq:MS} as a function of the end-to-end distance. The red line indicates the stall force of cohesin measured in the study, at 0.3 pN. The green line represents the force threshold at which CTCF proteins can efficiently block the loop extrusion process, at 0.14 pN. Below 0.14 pN, CTCFs fail to block loop extrusion. The tension range between 0.14 pN and 0.3 pN contains the loop limited regime discovered in our study.
{\bf Right:} DNA tension as a function of $L_{\rm end}$ and $L_A$, calculated using Eq.~\ref{eq:MS} with $L_c$ reduced by $L_A$. The black line indicates the transition between the tension limited and loop limited regimes in our study. The white regions represent forces below 0.14 pN or above 0.3 pN.
}
\end{figure}
\begin{align} 
\begin{split}
\label{eq:MS}
f_{}(L_{\rm end},L_c)=\frac{k_BT}{l_p} \Big(\frac{1}{4}\frac{1}{(1-L_{\rm end}/L_c)^2}-\frac{1}{4}+\frac{L_{\rm end}}{L_c}\Big).
\end{split}  
\end{align}

{\color{cyan}
The phase diagram in Fig. 3g of the main text has an intriguing connection to recent experimental findings on the tension-regulated control of loop extrusion by CTCF. Davidson et al.~\cite{davidson2023ctcf} discovered that CTCF functions as a DNA tension-dependent barrier, efficiently halting loop extrusion when DNA tension exceeds 0.14 pN. This finding suggests a regulatory mechanism for loop extrusion governed by DNA tension. Given that the maximum stalling force for cohesin-mediated loop extrusion observed in experiments is 0.3 pN, we find that our study aligns well with these results.
In Figure \ref{Fig:force_CTCF}  (left) in the SI, we show DNA tension calculated from Eq.\ref{eq:MS}, along with lines indicating 0.14 pN and 0.3 pN. The plot reveals an overlap between the loop limited regime found in our study and the effective force range for CTCF regulation of loop extrusion. As the DNA loop length increases due to loop extrusion, tension along the DNA is expected to accumulate. In Figure \ref{Fig:force_CTCF}  (right), we present DNA tension as a function of end-to-end distance $L_{\rm end}$ and DNA loop length $L_A$, clearly illustrating the overlap between the force regime for CTCF regulation and the transition line between tension limited and loop limited regimes in our study. 
}

\section{Probability of condensate formation}
We can compute the probability of the condensation using the obtained free energies. We first computed the probability density creating DNA-protein co-condensates as a function of $L_d$. For the scenario Fig.3a in the main text,
\begin{align} 
\begin{split}
\label{}
P_{a}(L_d) = \frac{e^{-\beta \Delta F_{a}(L_d)}}{\int_0^{L_{\rm max}} d L_d e^{-\beta \Delta F_{a}(L_d)}},
\end{split}  
\end{align}
where $L_{\rm max}=L_c-L_{\rm end}-L_A$. The probability density has either single or double peaks ($L_d=0$ and $L_d >0$) depending on the parameters. The probability of the condensate formation was determined as the fraction of the peak value located at $L_d >0$ to the sum of the peak values at $L_d=0$ and $L_d >0$. 




\section{Analysis of simulation data}
\subsection{Condensate detection}
For the identification of condensates within our simulation trajectories, we employed Density-based Spatial Clustering of Applications with Noise (DBSCAN~\cite{ester1996density}). DBSCAN operates based on two key parameters: the distance threshold $\epsilon$ and the minimum number of points required to constitute a dense region, denoted as minPts. In our analysis, we configured these parameters to $\epsilon=20 \ \mathrm{nm}$ and $\mathrm{minPts}=30$. By applying DBSCAN to the particle coordinates from the simulation trajectories, we define regions classified as condensates.

\subsection{Contact Maps}
To construct the contact map, we analyzed the last 1000 frames of the simulation, averaging the data across three distinct ensembles for each parameter set. A contact is defined as occurring when two particles are within $30 \mathrm{nm}$ of each other. The metric $I$ is calculated as the values in the contact map normalized by the total number of possible interactions ($1650^2$).


\bibliography{mybib}